\documentclass[conference]{IEEEtran}
\IEEEoverridecommandlockouts
\usepackage{cite}
\usepackage{amsmath,amssymb,amsfonts}
\usepackage{algorithmic}
\usepackage{graphicx}
\usepackage{textcomp}
\usepackage{xcolor}
\def\BibTeX{{\rm B\kern-.05em{\sc i\kern-.025em b}\kern-.08em
    T\kern-.1667em\lower.7ex\hbox{E}\kern-.125emX}}
\begin{document}

\title{DCM: A Developers Certification Model for Mobile Ecosystems
}

\author{\IEEEauthorblockN{Paulo Trezentos}
\IEEEauthorblockA{\textit{Aptoide S.A.}\\
Lisbon, Portugal \\
paulo.trezentos@aptoide.com}
\and
\IEEEauthorblockN{Ricardo Capote}
\IEEEauthorblockA{\textit{Aptoide S.A.}\\
Lisbon, Portugal \\
ricardo.capote@aptoide.com}
\and
\IEEEauthorblockN{Tiago Teodoro}
\IEEEauthorblockA{\textit{Aptoide S.A.}\\
Lisbon, Portugal \\
tiago.teodoro@aptoide.com}
\and
\IEEEauthorblockN{João Carneiro}
\IEEEauthorblockA{\textit{Aptoide S.A.}\\
Lisbon, Portugal \\
joao.carneiro@aptoide.com}
}

\maketitle

\begin{abstract}
This article introduces a distributed model of trust for app developers in Android and iOS mobile ecosystems. The model aims to allow the co-existence of multiple app stores and distribution channels while retaining a high level of safety for mobile device users and minimum changes to current mobile operating systems.

The Developers Certification Model (DCM) is a trust model for Android and iOS that aims to distinguish legit applications from security threats to user safeness by answering the question: “is the developer of this app trustable”?

It proposes security by design, where safety relies on a chain of trust mapping real-world levels of trust across organizations. For the technical implementation, DCM is heavily inspired by SSL/TLS certification protocol, as a proven model that has been working for over 30 years
\end{abstract}


\section{Introduction} \label{sec:introduction}
The mobile ecosystem includes 3 billion Android and 2 billion iOS-based devices. Those devices (phones, tablets, smartwatches,…) are used for entertainment, information, and communication.

The thriving mobile ecosystem relies heavily on the safety and trustworthiness of apps’ distribution.

However, there are strong incentives for bad actors to use apps and games to steal user data and perform hidden actions that may hurt the user.

In the model of one app store per operating system, Google and Apple implemented a centralised model when each company performed the curation and vetting the apps installed in the mobile devices. This was implemented through the app store or, in case of Google, later extended to Google Play Protect on-device screening for apps installed outside the store.

May one agree or not that the centralised model works for creating a trustworthy environment, it is becoming more obvious that the lack of competition in apps distribution may lead to worst business conditions to developers and less innovation in users’ experience.

Regulators and legislators across the world are proposing changes that is expected to have impact in the ecosystem, promoting competition across app stores.

Is it possible to have app store competition while maintaining the mobile user safety?

To answer to that question, the authors of this article propose a cascading trust model with cascade layers of vetting and different levels of trust.

There are no perfect models, but Internet SSL protocol exists since the mid 1990s in a comparable and similar environment. The trust in websites can be compared with trust of apps. The organisation claiming to be owner of the site can be compared with the developer claiming as the legit owner of the app.

This paper is structured as follows: in section \ref{sec:state-of-the-art}, we give an overview of SSL's main concepts and features supporting proof of identity. In section \ref{sec:dcm} we introduce the basic concepts of DCM model and in section \ref{sec:attack-vectors} we discuss possible attack vectors and how they are handled by DCM. Section \ref{sec:implementation} introduces suggestions for model implementation. Finally, section \ref{sec:conclusion} provides conclusions about the proposed model.

\section{State of the art} \label{sec:state-of-the-art}

\subsection{Secure Sockets Layer Protocol}

Developed in the mid 1990s by Netscape Communication Corporations, Secure Sockets Layer (SSL) is a cryptographic protocol that is used to establish secure communication between web browsers and servers. The protocol utilizes a combination of public key and symmetric key encryption techniques to establish a secure channel for transmitting private information. SSL has since been replaced by Transport Layer Security (TLS) protocol, a standard with its most current version defined in RFC 8446 \cite{rfc8446}, which uses the same basic principles as SSL providing improved security and it is widely adopted as the standard for secure communication \cite{10.1002/sec.1113}

\begin{figure*}
\centerline{\includegraphics{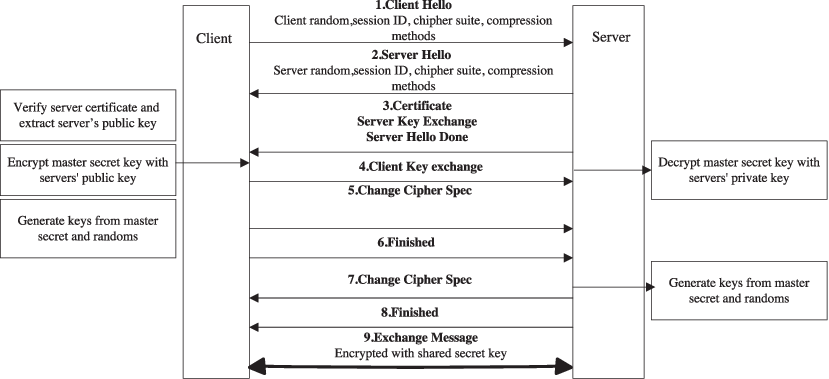}}
\caption{Steps involved in a full handshake\cite{10.1002/sec.1113}.}
\label{fig:ssl-handshake}
\end{figure*}

The SSL/TLS handshake is the process by which the client and server establish a secure connection. The handshake protocol steps are shown in figure \ref{fig:ssl-handshake}.

Once the SSL/TLS handshake is complete, the client and server can communicate securely using symmetric encryption with the session keys generated during the handshake.

The establishment of a secure connection in SSL/TLS relies on digital certificates, which are issued by Certification Authorities (CAs) or other trusted entities. These certificates serve as proof of identity for the server and allow web browsers to verify the identity of the intended server and establish a secure connection.

The certificate includes typically includes a set of fields Common Name (CN), Organization (O), Organizational Unit (OU), City/Locality (L), State/County/Region (S), Country(C) and Email Address \cite{rfc8446}.

The standard allows including additional fields through a Certificate Extension for supporting information and applications of given communities \cite{rfc8446}.

When verifying the validity of a certificate, the browser’s security checks include:

\begin{itemize}
\item Verify the domain against what is listed in the certificate;

\item Verify if issuing CA is part of the list of trusted CA;

\item Certificate expiration date;

\item Status valid vs revoked (CRL or OCSP);

\item Digital signature of the certificate (CA validation) \cite{rfc5280}.
\end{itemize}

\subsection{Certification Authorities}

Certification authorities are trusted entities that issue digital certificates to verify the identity of servers and other entities on the internet. The role of CAs in SSL/TLS is critical for establishing trust between web browsers and servers, as they ensure that the certificates are issued only to legitimate entities and that the certificates themselves are not compromised.

A CA hierarchy exists consisting of a root CA, intermediate CA, and end-entity CA. The root CA is the top-level CA in the hierarchy, which is trusted by all web browsers and operating systems. The root CA issues digital certificates to intermediate CAs, which in turn issue certificates to end-entity CAs or directly to end-entities. The intermediate CAs are trusted by the root CA, and the end-entity CAs are trusted by the intermediate CAs.

The CA hierarchy plays an important role in determining the trustworthiness of the CA in issuing the digital certificates.

\subsection{Online Certificate Status Protocol} \label{subsec:ocsp}

While the Certificate Revocation List (CRL) method relies on the periodic download of a list of revoked certificates \cite{rfc6962}, the Online Certificate Status Protocol provides a more efficient and timely method for checking the revocation status of digital certificates.

The Online Certificate Status Protocol (OCSP) is a protocol that allows applications to verify the status of certificates. It is a proposed standard described in RFC8980\cite{rfc6960}.

To verify the validity of a certificate, the application sends a query to the OCSP responder, which is a server maintained by the certificate issuer or a trusted third-party provider. The OCSP responder checks its status and responds with one of three statuses: good, unknown, or revoked.

\subsection{Certificate Transparency System} \label{subsec:ct}

Certificate Transparency (CT) is an internet security standard that logs all issued digital certificates for monitoring and auditing purposes, enabling validation of their authenticity and detection of any malicious activities within the digital certificate ecosystem\cite{rfc6962}.

The CT system works by requiring certification authorities to publicly log all SSL/TLS certificates that they issue in a publicly accessible log. This log is monitored by various entities, including browser

vendors, security researchers, and other certification authorities, to detect potentially fraudulent or malicious certificates.

If a fraudulent or malicious certificate is detected, the CT system allows for the revocation of the certificate by the certification authority that issued it. This enables other entities to be alerted to the potential threat and take appropriate action to mitigate the risk.

The SSL/TLS protocol is an established standard that provides technology to support identity authentication and secure communication. The next section describes how the TLS can be used as a distributed developer trust framework. 

\subsection{Malware Information Sharing Project (MISP)} \label{subsec:misp}

Threat Intelligence platforms are tools used by Threat Researchers and Organizations to share Indicators of Compromises (IOC), Security Exploits or Malware Samples, allowing stakeholders to share information and respond in real-time.

The MISP - Open Source Threat Intelligence and Sharing Platform \cite{10.1145/2994539.2994542} is an open source project actively developed since 2011 by organisations including the Belgian Defense and NATO and promoted among CERT organization.

MISP shares and stores indicators of compromises of targeted attacks, enabling organizations to share information such as threat intelligence, indicators, threat actor information or any kind of threat which can be structured in MISP. MISP community benefits from the collaborative knowledge about existing malware or threats, improving the counter-measures used against targeted attacks and set-up preventive actions and detection. The goals of MISP are: facilitate the storage of technical and non-technical information about seen malware and attacks, create automatically relations between malware and their attributes, store data in a structured format (allowing automated use of the database to feed detection systems or forensic tools), share malware and threat attributes with other parties and trust-groups and improve malware detection and reversing to promote information exchange among organizations (e.g. avoiding duplicate works).

\section{Developers Certification Model (DCM)} \label{sec:dcm}

The security of a software application has been the object of various research projects and industry projects. From software applications to more recent mobile apps, anti-virus companies and app stores developed algorithms and frameworks to evaluate how trustable is an app.

Those security frameworks are based in static code analysis, code fingerprinting and run-time tracing to identify possible threats to user safeness. The DVM model doesn’t aim to replace such efforts but rather complement it.

Current security efforts developer by app stores aim to answer the question: “is this app safe?”, while DCM answers the question “Is this developer trustworthy?”.

The model is built in three basic principles:

\begin{itemize}
    \item \textbf{Implementation feasibility}: it needs to be able to be implemented today, having existent mobile ecosystems and reasonable resources. The implementation should be simple and easy to execute by the different stakeholders;

    \item \textbf{Security by design}: the implementation of security should map reality and not ambition to shape a new reality that fits the model;

    \item \textbf{Distributed and decentralised}: mobile devices are heterogenous working over different geographies, different networks and different protocols. A trust model for such a diverse platform must be distributed and decentralised to handle the scale and the potential points of failure.
\end{itemize}

The implementation of DCM model certifies a developer in one of 4 trust levels:

\begin{itemize}
    \item \textbf{Trusted}: the developer has provided information and has a track record that allows to classify her as trustable until contrary information. A trusted developer is some person or organisation which was proven legitimacy through: a) proof-of-incorporation, personal identification, address verification and b) has an app distributed for more than 2 years with more than 10.000 installs. The exact requirements should be defined later in the implementation phase;

    \item \textbf{Unknown}: not enough information is available, or the developer didn’t meet with defined requirements to be classified as trust but has no information of threat or malicious behavior associated;

    \item \textbf{Warning}: a specific pattern or behavior that may result in a threat to the user was identified. The behavior of certain users can be acceptable while for others not. This can be for instance the case of invasive advertisement or mishandling of user information. Power users with enough information can decide to take the risks.

    \item \textbf{Critical}: there was a serious issue identified in one or more apps of the developer and the app should not be installed in any case, posing a clear danger for the user.
\end{itemize}

When issuing a certificate for a given developer, the CA evaluates and assigns a trust level to the developer. The trust level is then included as an additional attribute to the certificate, as a Certificate Extension described in \cite{rfc8446}.

\begin{figure}[htbp]
\centerline{\includegraphics[width = 0.4 \textwidth]{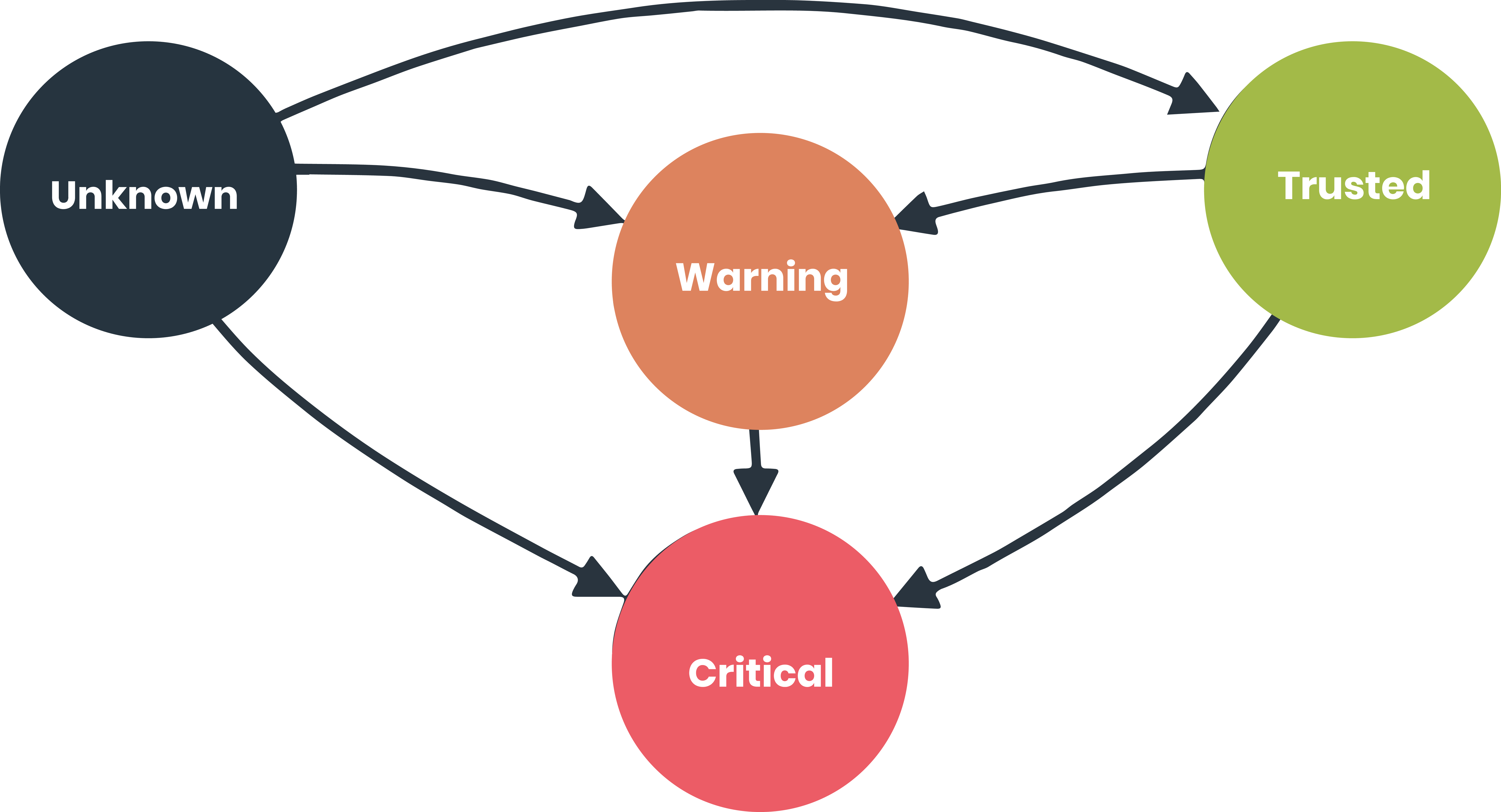}}
\caption{Diagram of the trust level changes.}
\label{fig:trust-level-changes}
\end{figure}

Figure \ref{fig:trust-level-changes} depicts the changes in trust level that can be expected to occur over time.

Changes in trust level may arise during the reevaluation of a developer by the Certification Authority (CA) responsible for issuing a certificate, when the CA concludes that the trust level has shifted since

the last assessment. Upon a change in the trust level, the previously issued certificate is revoked by the CA, and a new certificate must be issued, reflecting the updated trust level.

A competitive landscape implicates that we may have dozens of app stores with the security mechanisms to detect dangerous apps, with their own information systems and business rules. DCM promotes the enough level of cooperation among app stores and ecosystem actors leaving space for differentiation and strategies.

The following entities participate in DCM with different roles:

\begin{itemize}
    \item \textbf{Root Certification Authority (RCA)}: a root certification authority is an organisation that is responsible to certify the legitimacy of intermediate certification authorities ICA). Its role is to guarantee that candidate organizations for ICA are idoneous and have the human and technical resources to certify developers and accomplish the mission. The Root certification authorities would Google, Apple, European Commission, and Federal Communications Commission (FCC). Technically, RCAs issue a digital certificate that enables ICA to conduct their business.

    \item \textbf{Intermediate Certification Authority (ICA)}: an ICA is an entity - most likely an App Store - that certificates that a developer meets the requirements of DCM as “Trustable”. ICA has also the power to classify the developer in the Warning or Critical trust level if threats are detected in their apps.

    \item \textbf{Mobile Operating Systems Installation Managers (MOSIM)}: installation managers are responsible at operating system level to allow or not the installation of app packages. In DCM, MOSIM in Android and iOS allow the smooth installation of app packages coming from developers with “Trustable” level, without friction. In the case of apps from developers with state “Unknown” or “Warning” a message may be displayed to the user, as well as the enough information for her to be able to decide on the risk to move forward with the installation of the app. Apps from developers classified as “Critical” are not installable. If an Internet connection is available, MOSIM checks revocation lists at install time for revocation in the trust level (see OCSP - Online Certificate Status Protocol section \ref{subsec:ocsp}).

    \item \textbf{App Stores}: app stores role is twofold. On one hand, app stored screen and vet apps, issuing certificates as Intermediate Certification Authority (ICA). On the other hand, app stores are the hub for apps and games distribution and the interface with the user. In the later role, app stores can decide which ICA certification they accept, and which trusted level apps they want to list. They have the freedom to choose which certified apps to accept and existing screening processes for accepted apps are likely to be maintained.

    \item \textbf{Developers}: developers develop apps and games, seeking certification for those in one ICA. After certification completion, developers receive a digital certification that can be used to sign any app, in every version.

    \item \textbf{Apps}: apps and games are mobile software for entertainment, leisure, information or communication signed with the digital certificate of the developer. The level of trust is the same across all apps of a developer.
\end{itemize}

\begin{figure*}[!htbp]
\centerline{\includegraphics[width = 0.8 \textwidth]{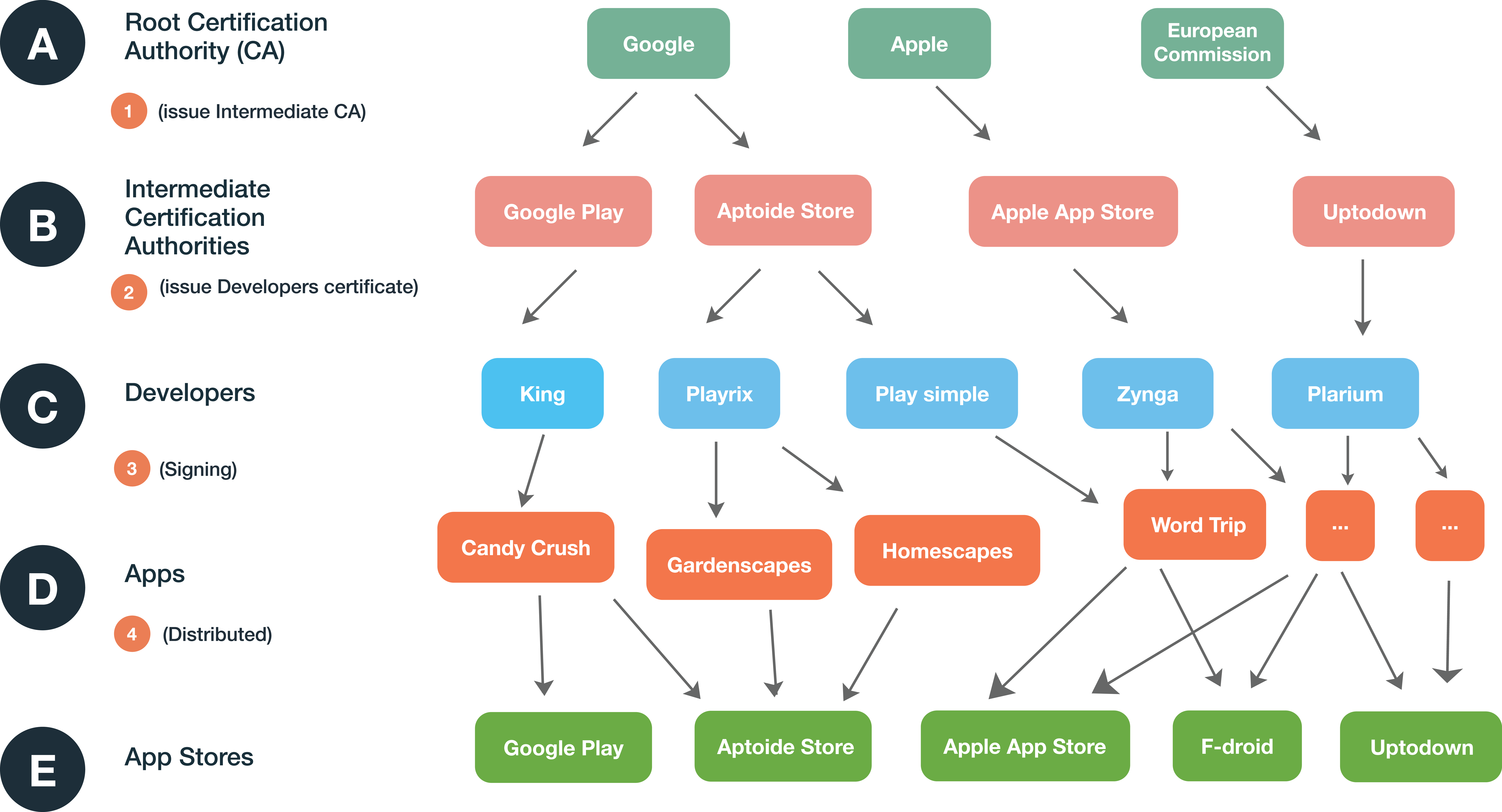}}
\caption{Stakeholders and interactions within DCM.}
\label{fig:dcm-stakeholders}
\end{figure*}

The model in figure \ref{fig:dcm-stakeholders} depicts the above concepts.

\subsection{Validation Steps}

When installing a mobile application, the package manager (MOSIM) should verify the developers’ certification status. DCM proposes a validation comprised of the following six steps:

\begin{enumerate}
    \item Developer’s Signature: The package manager checks that the developer signature of the app matches the developer signature listed in the SSL/TLS certificate.
    
    \item Certificate authority (CA): The package manager checks that the SSL/TLS certificate is issued by a trusted certificate authority. The same as web browsers, each package manager has a list of trusted CAs, and the certificate presented by the server must be issued by one of these trusted authorities.
    
    \item Expiration date: The package manager checks that the SSL/TLS certificate has not expired.
    
    \item Revocation status: The package manager checks that the SSL/TLS certificate has not been revoked by the certificate authority. The revocation status of the certificate can be confirmed by looking up the certificate revocation list (CRL) or using the online certificate status protocol (OCSP).
    
    \item Digital signature: The package manager checks the digital signature on the SSL/TLS certificate. The certificate includes a digital signature that is verified by the package manager using the public key of the certificate authority that issued the certificate.
    
    \item Trust level: The package manager checks the trust level attribute listed in the certificate to verify if the developer has obtained Trusted score from the issuing CA.
\end{enumerate}

If all steps are successfully validated, the package manager can confirm trustworthiness of the developer and proceed with the installation of the mobile app. If any of the validation steps fails, the package manager may deny installation or prompt the user for further action.

\section{DCM attack vectors} \label{sec:attack-vectors}

Bad actors will try to explore the vulnerabilities of DCM to push malware apps and games to user mobile phone. In this section, we explore some of the potential attack vectors and how DCM is prepared to handle such threats.

\subsection{The weakest link attack}

A chain is no stronger than its weakest link.

Having multiple certification authorities, a bad actor will always try to have her certificate issued by the certification authority with the poorest vetting and screening process. After the certificate issued, the bad actor will try to publish into the app store with the poorest anti-malware detection platform exploring the weakest links in DCM ecosystem. Regarding the issue of the issuing of the certificate, the Certificate Transparency system (introduced in section \ref{subsec:ct}) gives total visibility of certificates issued by each CA, enabling an early detection of CAs that do not comply with the rules to approve developers. A CA that doesn’t implement the criteria defined for a developer to be certified, can see its Intermediate certificate revoked and be no longer part of DCM CA chain.

On the other hand, Android app stores can further improve the collaboration among malware threats by using a platform like MISP, introduced in section \ref{subsec:misp}. Using such a collaboration platform, the safeness of apps submitted to app stores would be leveled up and submitting the app to the app store with poorest anti-malware detection would not work anymore, as information about the potential malware (such as fingerprinting of dangerous procedures and malware links) would propagate quickly across Android app stores.

Together, the Certificate Transparency system and MISP-like platform would make weakest link attacks harder and with no more impact than the existent in Android ecosystem.

\subsection{Dormant developer attack}

In the dormant developer attack, the bad actor developer receives an “Unknown”-level certificate for an innocuous app, waits for the enough time / enough downloads to become “Trusted” and then releases an update version of the app with malware code.

DCM handles this case in two steps. In the first step, the updated version will be detected by the app store upon submission and before release. In the case the app store doesn’t detect the malware immediately, through the MISP platform the app would be screened by other app stores and a malware-warning notice issued. After receiving the malware information, the app store automatically removes the malware app.

In a second step, the certification authority that issued the bad-actor developer certificate would revoke it, eliminating the risk of that version being installed outside the Android app stores network, for instance through side-loading.

\section{Implementation} \label{sec:implementation}

In this section, we first review the existing implementations adopted by each of the major mobile systems: Android (section \ref{subsec:soa-android}) and IOS (section \ref{subsec:soa-ios}). Based on the existing methodology, section \ref{subsec:dcm-implementation} proposes the implementation details for adopting the proposed DCM. \\

Both implementations share a similar sequential logic, which includes the following steps:

\begin{enumerate}
    \item Obtain a certificate with a private/public key pair.

    \item Sign the app that the developer intends to publish.

    \item The operating system verifies the app's signature and files integrity before allowing its installation.
\end{enumerate}

\subsection{Android} \label{subsec:soa-android}

\textbf{STEP 1 - Obtaining a certificate}

Google’s Play Store current process of obtaining a certificate is the following:

\begin{enumerate}
    \item The developer signs up for a Google Play Console account.

    \item The developer is required to create a digital identity that is used to sign their app.

    \item Google issues the digital identity by generating a public-private key pair.

    \item The developer uploads the app to be signed with their private key, creating a digital signature that can be verified using the public key.
\end{enumerate}

\textbf{STEP 2 - Signing the app}

Google Play uses the app signing to verify the authenticity of the app and to ensure the integrity of the app's signature. This helps to protect users from downloading and installing malicious or counterfeit apps from the Google Play store.

The process of signing an .apk can be described in the following sequential steps:

\begin{enumerate}
    \item The developer builds the APK file from the source code and resources of the app.

    \item A digest of the APK file is generated, which is a hash of the file's contents.

    \item The digest is signed with the private key using a digital signature algorithm, which produces a signature. The signature includes a hash of the APK file, and the private key of the developer.

    \item The signature is added to the APK file, along with the certificate and the public key that corresponds to the private key.

    \item The signed APK file is verified to ensure that the signature is valid and matches the contents of the APK file.
\end{enumerate}

\textbf{STEP 3 - Verify the signature}

\begin{enumerate}
    \item Obtain the app's digital signature. The digital signature is stored in /META-INF folder, which contains metadata about the app, including the files:

    \begin{itemize}
        \item MANIFEST.MF - which lists all the files included in the APK and their corresponding SHA-1 checksums. This file is used to verify the integrity of the APK during the installation process.
    
        \item CERT.RSA - which is the Android certificate that contains the public key to verify the app’s signature.
    
        \item CERT.SF - which contains a list of all files in the APK and their digital signatures.
    \end{itemize}

    \item Check the app certificate to ensure that it is valid.

    \item Verify the app's digital signature. The operating system verifies the app's digital signature. This includes checking:

    \begin{itemize}
        \item if certificate is not expired;

        \item that the package name is the one specified in the app's manifest file, including verifying the APK file's contents, checksums, and file sizes;

        \item that the file signature of the app matches the signature stored in the APK's META-INF/ folder.
    \end{itemize}

    \item Allow or prevent installation:

    \begin{itemize}
        \item If the app's signature is valid, the operating system allows the app to be installed and run on the Android device. If the signature is invalid, the operating system will prevent the app from being installed.
    \end{itemize}
    
\end{enumerate}

\subsection{iOS} \label{subsec:soa-ios}

\textbf{STEP 1 – Obtaining a certificate}

Apple manages developers' certificates through the Apple Developer Program. The program employs multiple intermediate certificate authorities (CA's) that are subordinate to the Apple Root CA. Some of these intermediate CA's may be controlled by Apple, while others may not be. Nevertheless, Apple retains control over the revocation of any developer certificate since it controls the only root CA.

Apple’s current process of obtaining a certificate is the following:
\begin{enumerate}
    \item The developer enrolls in the Apple Developer Program by paying an annual fee.

    \item The developer generates a certificate signing request (CSR) on their local machine.

    \item The developer submits the CSR to Apple through the Apple Developer website.

    \item Apple verifies the developer's identity and issues a developer certificate.

    \item Apple stores the developer's private key on a secure server and provides the developer with a signed public key certificate.

    \item The developer downloads and installs the public key certificate on their local machine.

    \item The developer uses the certificate to sign their applications or updates.
\end{enumerate}

\textbf{STEP 2 – Signing the app}

Apple's current signing process is called “codesign”. It is used to verify the authenticity of the developer who publishes an app through the App Store and to ensure that the contents of the app were signed by the same developer.

The process of codesigning an “.ipa” (iOS package) can be described in three sequential steps:

\begin{enumerate}
    \item Recursive signing of nested code:

    \begin{itemize}
        \item With the developer’s private key sign all the helpers, tools, libraries, frameworks, and other components that the app relies on and are bundled with the app.
        \item These digital signatures are stored in ''CodeSignature/CodeResources''. 
    \end{itemize}

    \item Signing of Mach-O Executables:

    \begin{itemize}
        \item The signing software applies individual signatures to each architectural component of a universal binary that represents the main executable of the app.

        \item These signatures are independent, and usually only the native architecture on the end-user's system is verified.

        \item To apply the signature, the codesign utility adds the signature directly to the executable file (CMS signature).
    \end{itemize}

    \item Signing of Resources:

    \begin{itemize}
        \item Everything in an application bundle that is not explicit code (either nested code bundles or the main executable) is a resource and is signed.

        \item The resource files themselves are not modified as result of signing. “codesign” places the digital signatures corresponding to all the application bundle’s non-code files in a special plist file within the bundle, namely \_CodeSignature/CodeResources.
    \end{itemize}
\end{enumerate}

\textbf{STEP 3 – Verify the signature}

\begin{enumerate}
    \item Obtain the app's digital signature:

    \begin{itemize}
        \item This signature is embedded in the executable’s binary and contains information that identifies the app and its developer.
    \end{itemize}

    \item Check the root certificate authority (CA):

    \begin{itemize}
        \item The operating system checks the app's signature against the root CA that issued the certificate used to sign the app. In Apple's case, this is the Apple Root CA and its public key is stored in the OS.
    \end{itemize}

    \item Verify the intermediate CAs:

    \begin{itemize}
        \item If the root CA is trusted, the operating system proceeds to verify the intermediate CAs that signed the app's certificate. These intermediate CAs may or may not be controlled by Apple, but Apple keeps control of revoking any developer certificate since it controls the only root CA.
    \end{itemize}

    \item Verify the app's digital signature:

    \begin{itemize}
        \item Once the chain of trust is established, the operating system verifies the app's digital signature. This includes checking the signature of the app's main executable, any nested code, and all resources within the app bundle.
    \end{itemize}

    \item Verify individual components of the app:

    \begin{itemize}
        \item During the verification process, the operating system checks the digital signature of each individual architectural component of the app's universal binary.
    \end{itemize}

    \item Allow or prevent installation:

    \begin{itemize}
        \item If the app's signature is valid and can be traced back to a trusted root CA, the operating system allows the app to be installed and run on the iOS device. If the signature is invalid or cannot be traced back to a trusted root CA, the operating system will prevent the app from being installed.
    \end{itemize}
\end{enumerate}

\subsection{Proposed implementation} \label{subsec:dcm-implementation}

To bridge the gap between the proposed Developer Certification Model and the existing processes of Apple and Google, the suggested implementation aims to maintain, almost totally, the current signing and verification processes, as well as the file formats (.apk and .ipa). The distribution of certificates would differ by creating a chain of CAs that certify developers. Additionally, the certificate status would classify a developer's certificate and their signed applications as unknown, warning, critical, or trusted, depending on their level of trustworthiness.

\begin{figure}[htbp]
\centerline{\includegraphics[width = 0.5 \textwidth]{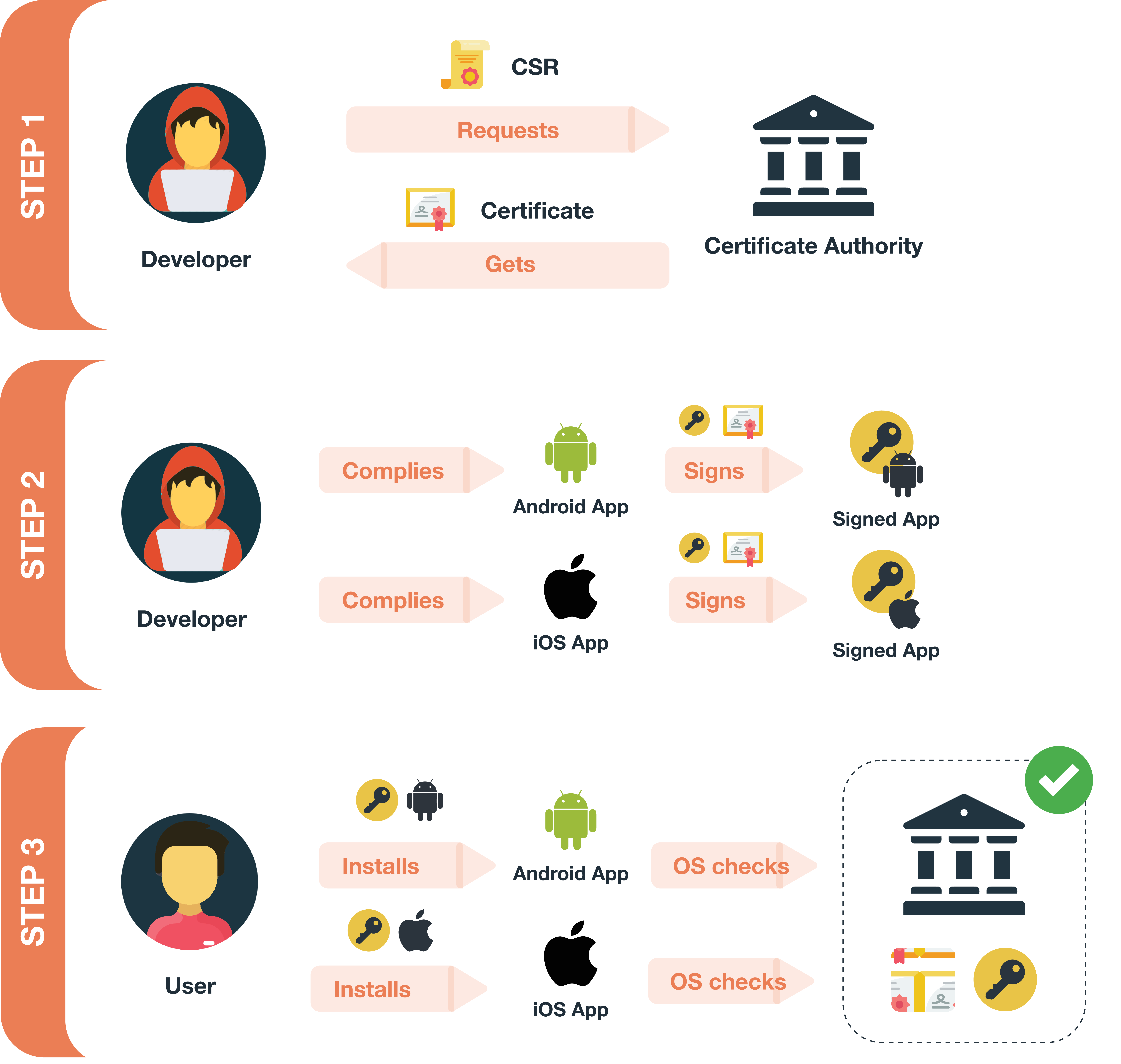}}
\caption{Flow of the Developer Certification Model.}
\label{fig:dcm-flow}
\end{figure}

The proposed implementation is shown in figure \ref{fig:dcm-flow} and described by the same three steps as before:

\textbf{STEP 1 – Obtaining a certificate}

To ensure consistency and convenience for developers, the proposed model advocates for a unified process of obtaining a certificate across different operating systems. With this approach, developers will have the ability to use the same certificate to sign their apps in both iOS and Android.

To obtain a certificate, a developer must request it from an intermediate CA that has a properly signed certificate issued by one of the root CAs. Once the intermediate CA has verified the developer's identity, they will provide the developer with the certificate that authenticates the developer.

The developer generates the private key that will be used to sign their apps together with the CA’s certificate.

Apple already employs a similar trust chain, but it is currently centralized, with Apple being the only root CA and holding the authority to revoke any developer's certificate. Our proposal, on the other hand, seeks to decentralize this process.

\textbf{STEP 2 – Signing the app}

Our approach is to maintain a similar signing process to that of Apple and Google's current processes, apart from adopting the mutual certificate format, which should not require significant changes.

\textbf{STEP 3 – Verify the signature}

During the verification process before installing an app, the Package Manager (MOSIM) should not only verify the app's signature but also check the validity of the developer's certificate. This includes:

\begin{itemize}
    \item verifying the chain of certification authorities (CA) of the app certificate to ensure it is valid and that it can be traced back to a trusted certificate authority (by using a dictionary of all known public keys, similar to how browsers work).

    \item check if the certificate is still valid and has not been revoked (via OCSP protocol) or expired.

    \item check that the package name and file signature of the app matches the signature stored in the app metadata.

    \item check the app file's contents, checksums, and file sizes.
\end{itemize}

Based on the verification results, the system should take appropriate actions, such as preventing the installation of apps with unverified signatures or flagging apps with certificates indicating low trust levels to prompt user attention.

\section{Conclusion} \label{sec:conclusion}

The foreseen possibility of allowing side load / installation of mobile apps by third parties in mobile devices poses opportunities and threats. The opportunities are related with healthy competition among players and corresponding innovation. The challenge is how can we block bad actors while allowing a trustworthy installation process open for recognized entities?

The solution for this challenge will never be optimal as today’s security in mobile devices is not perfect either. Nevertheless, Developers Certification Model proposes a framework that can be implemented with minimal changes in Android and iOS and solves the challenge by answering the question: “is this developer trustworthy?”. To answer the question in an efficient way, DCM proposes a model that merges existing SSL protocol for Web with current package managers implementation.

A reference implementation in Android and iOS was developed – and is publicly available- to validate the feasibility of the proposals in this article.

Finally, we believe that this article is just the initial step of a broader discussion among legislators, regulators, gatekeepers, app stores and developers toward a certification model that can unleash a safe environment for a new phase of the mobile economy.

\nocite{*}
\bibliographystyle{IEEEtran}
\bibliography{dcm}

\end{document}